\documentclass[aps,preprint,prd,showpacs,nofootinbib]{revtex4}
\usepackage{amsmath}
\usepackage{graphicx}
\usepackage{dcolumn}
\usepackage{bm}
\usepackage{amssymb}
\usepackage{latexsym}

\def\be{\begin{equation}}
\def\ee{\end{equation}}
\def\ba{\begin{eqnarray}}
\def\ea{\end{eqnarray}}

\begin{document}

\title{ Is the island universe model consistent with observations?
}

\author{Yun-Song Piao$^{a,b,c}$}
\affiliation{${}^a$Institute of High Energy Physics, Chinese
Academy of Science, P.O. Box 918-4, Beijing 100039, China}
\affiliation{${}^b$Interdisciplinary Center of Theoretical
Studies, Chinese Academy of Sciences, P.O. Box 2735, Beijing
100080, China} \affiliation{${}^c$College of Physical Sciences,
Graduate School of Chinese Academy of Sciences, YuQuan Road 19A,
Beijing 100049, China}

\begin{abstract}

We study the island universe model, in which initially the
universe is in a cosmological constant sea, then the local quantum
fluctuations violating the null energy condition create the
islands of matter, some of which might corresponds to our
observable universe. We examine the possibility that the island
universe model is regarded as an alternative scenario of the
origin of observable universe.

\end{abstract}

\pacs{98.80.Cq} \maketitle

Recent observations of the cosmic microwave background (CMB) have
implied that the inflation \cite{Guth, LAS} is a very consistent
cosmological scenario. The inflation stage took place at the
earlier moments of the universe, which superluminally stretched a
tiny patch to become our observable universe today, and ended with
a period of reheating that bring the universe back to the usual
FRW evolution. The quantum fluctuations in the inflation period
are drew out the horizon and, when reentering into the horizon
during radiation/matter domination after the end of inflation, are
responsible for the formation of cosmological structure.

However, the inflation has still its own problem of fine tuning,
in addition it is not also able to address the initial singularity
problem, the cosmological constant problem, super Planck
fluctuations problem, as was described in Ref. \cite{B}. Thus it
seems desirable to search for an alternative to the inflation
model. Recently, Dutta and Vachaspatian have proposed a model
\cite{DV}, in which initially the universe is in a cosmological
constant ($\Lambda$) sea, then the local quantum fluctuations
violating the null energy condition (NEC) \cite{Wini, Vach} create
the islands of matter, some of which might corresponds to our
observable universe. The headstream of this model in some sense
may be backward to the eternal inflation \cite{Vilenkin, Linde},
and especially the recycling universe proposed by Garriga and
Vilenkin \cite{GV}, in which the thermalised regions
asymptotically approaching dS spacetime may be fluctuated and
recycled back to the false vacuum and then the nucleated false
vacuum region will serve as a seed for a new (eternally) inflating
domain. The difference of island universe is that after the NEC
violating fluctuation is over, instead of the inflation followed,
the thermalisation occurs instantly and the radiation fills the
volume rapidly. Though the island universe model brings us an
interesting environment, people are still able to doubt whether
the island universe model is consistent with our observations.


Whether we can live in ``Islands"? We firstly attempt to
phenomenally (semiclassically) understand it, following closely
Ref. \cite{PZ}. We set $8\pi/m_p^2 =1$ and work with the parameter
$\epsilon = -{\dot h}/h^2$, where $h\equiv {\dot a}/a$ is the
local Hubble constant. The ``local" means here that the
quantities, such as $a$ and $h$, only character the value of the
NEC violating region. In the island universe model, the scale of
the NEC violating region is required to be larger than that of
present horizon, which has been clearly stated by Dutta and
Vachaspatian \cite{DV}, and is consistent with Refs. \cite{FG,
Linde92}, see also Refs. \cite{FDP, FGG, VT, BTV} for relevant
discussions. This result provides the initial value of local
evolution of $a$ and $h$. The $\epsilon$ can be rewritten as
$\epsilon \simeq {1\over h \Delta t}{\Delta h\over h}$, thus in
some sense $\epsilon$ actually describes the change of $h$ in unit
of Hubble time. During the NEC violating fluctuation, ${\dot
h}>0$, thus $\epsilon <0$ can be deduced. The time of this
fluctuation can be given by \be T=\int dt = -\int_i^e {dh \over
\epsilon h^2} \simeq {1\over |\epsilon| h_i} , \label{indt}\ee
where the subscript i and e denote the initial and end value of
the NEC violating fluctuation respectively. In this model $h_i^2
\equiv h_0^2\sim \Lambda$, where
the subscript 0
denotes the present value.

The reasonable and simplest selection for the local evolution of
scale factor $a(t)$, due to ${\dot h} >0 $, is \cite{PZ, PZ2} $
a(t) \sim (-t)^{n}$, where $t$ is from $-\infty$ to $0_-$, and $n$
is a negative constant. We have $h=n/t$, and thus $\epsilon= 1/n$.
To make $T \rightarrow 0$ in which the NEC violating fluctuate can
be so strong as to be able to create the islands of our observable
universe \cite{DV}, from (\ref{indt}), $|\epsilon | \rightarrow
\infty$ is required, which results in $n\rightarrow 0_-$. Thus
though the change of $h$ is large, the expanding proportion of the
scale factor is very small. Further in some sense before and after
the fluctuation $a$ is nearly unchanged, which can be also seen
from following discussions.

In the conformal time $d\eta =dt/a$, {\it i.e.} $-\eta\sim (-
t)^{-n+1}$, we obtain \be a(\eta) \sim (-\eta)^{{n\over
1-n}}\equiv (-\eta)^{1\over \epsilon -1} \label{aeta}\ee \be h
={a^\prime \over a^2}= {1\over (\epsilon -1) a \eta},
\label{htau}\ee where the prime denotes the derivative with
respect to $\eta$. The perturbations leaving the horizon during
the NEC violating fluctuation might reenter the horizon during
radiation/matter domination after the end of fluctuation, which
are responsible for the structure formation of our observable
universe. The efolding number which measures the quantity of
perturbation leaving the horizon during the NEC violating
fluctuation can be defined as \be {\cal N}_{ei} \equiv \ln{({a_e
h_e\over a_i h_i})} .\label{caln}\ee From (\ref{aeta}) and
(\ref{htau}), we obtain \be a\sim \left({1\over (1-\epsilon) a
h}\right)^{1\over \epsilon -1}\label{ah}\ee Thus for the constant
$\epsilon$, we have \be {a_e\over a_i}=({a_i h_i\over a_e
h_e})^{1\over \epsilon -1}= e^{{{\cal N}\over 1-\epsilon}} .\ee We
can see that for the negative enough $\epsilon$, the change
$\Delta a/a=(a_e-a_i)/a_i\simeq {\cal N}/(1-\epsilon)$ of $a$ will
be very small.

\begin{figure}[t]
\begin{center}
\includegraphics[width=6cm]{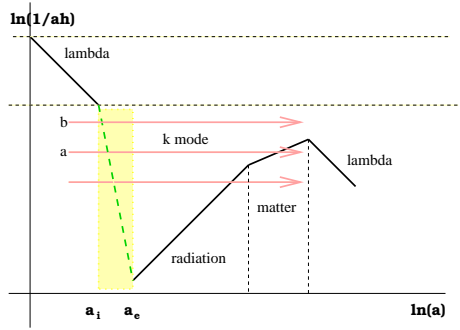}
\caption{ The sketch of evolution of local $\ln{(1/ah)}$ with
respect to the scale factor $\ln{a}$, in which the yellow region
denotes the NEC violating fluctuation and the red lines denote the
evolution of primordial perturbation modes. The perturbation modes
exit the Hubble horizon during the NEC violating fluctuation and
then reenter the horizon during the radiation/matter domination at
late time. }
\end{center}
\end{figure}

\begin{figure}[t]
\begin{center}
\includegraphics[width=6cm]{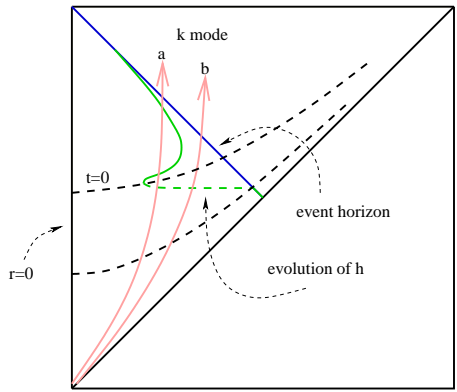}
\caption{The Penrose diagram of creation and evolution of island
universe, in which the blue line is the event horizon of $\Lambda$
sea, the green line and red lines denote the evolution of Hubble
scale and primordial perturbation modes respectively. The region
between black dashed lines denotes the NEC violating fluctuation.
The dashed line of $t =0$ is the thermalisation surface, after
which the NEC violating fluctuation is over, the local island
universe starts the FRW evolution, and with the lapsing of time
finally will return to the blue $\Lambda$ sea again. }
\end{center}
\end{figure}

Taking the logarithm in both sides of (\ref{ah}), we obtain \be
\ln{({1\over ah})}=(\epsilon -1)\ln{a}\equiv ({1-n\over n})\ln{a}
.\label{ahi}\ee We plot Fig.1 to further illustrate the island
universe model, in which $\ln{(1/ah)}$ is regarded as the function
of $\ln{a}$. In the NEC violating region, $n\rightarrow 0_-$, thus
the slop $(1-n)/n\rightarrow -\infty$, which makes the evolution
of $\ln{(1/ah)}$ during the fluctuation correspond to a nearly
vertical line in Fig.1. The (\ref{ahi}) can be also shown and
actually applied for the expansions with arbitrary constant $n$.
Thus during the $\Lambda$, radiation and matter domination, we
have the slope $-1$, $1$ and $1/2$ respectively, which have been
reflected in the Fig.1.

The NEC violating fluctuations in the $\Lambda$ sea with the
observable value of cosmological constant create some thermalised
matter islands, which subsequently evolute as the usual FRW
universe. The radiation and matter will be diluted with the
expansion of island and eventually this part of volume will return
to the $\Lambda$ sea again. The total evolution may be depicted in
the Penrose diagram of Fig.2. We can see that there are always
some perturbations (b) never reentering and remaining outside the
Hubble scale after their leaving from it during the NEC violating
fluctuation, which means that a part of island is permanently
inaccessible to any given observer in the island. This may be
reanalysed as follows. The wavelength of perturbations grows with
the scale factor $\sim a$, thus for the perturbations with the
present horizon scale, we have $a_0 h_0 = a_c h_c$, where the
subscript c denotes the value at the time when this perturbation
leaves the horizon during the NEC violating fluctuation. We define
\be {\cal N}_{ec} \equiv \ln({a_e h_e \over a_c h_c})=\ln({a_e h_e
\over a_0 h_0}) \ee as the efolding number when the perturbation
with the present horizon scale leaves the horizon during the NEC
violating fluctuation, and from (\ref{caln}), we have \be {\cal
N}_{ei}-{\cal N}_{ec}=\ln({a_0 h_0\over a_i h_i}). \ee Thus for
the island universe model, in which $h_i =h_0$ and $a_0\gg a_i\geq
1/h_i$, we obtain ${\cal N}_{ei}>{\cal N}_{ec}$. This implies that
as long as the islands created are suitable for our existence, the
efolding number required to solve the horizon problem of FRW
universe may be always enough, independent of the energy scale of
thermalisation.

Then we discuss the the primordial perturbations from the island
universe model. Dutta and Vachaspatian \cite{DV} have shown that
the spectrum from the quantum fields other than the NEC violating
field can be scale invariant, but the amplitude of perturbation is
too small to seed the structure of observable universe. However,
they in their calculations regarded the NEC violating region as an
instant link between the $\Lambda$ sea and FRW evolution, which to
some extent leads that the effect of NEC violating region on the
perturbation spectrums is lose. Thus we here will focus on the NEC
violating region.

To simplification, we adopt the work hypothesis in which the NEC
violating fluctuation behaves like the phantom energy. The
simplest implementation of phantom energy is a scalar field with
the reverse sign in its dynamical term, in which the NEC $\rho +p
=-{\dot \varphi}^2 <0 $ is violated. To make $\rho
>0$, the potential ${\cal V}(\varphi)> {\dot \varphi}^2/2$ of phantom field is also required.
Thus in this case
the local evolution of $h$ can be written
as $ h^2 \sim -{\dot \varphi}^2/2 +{\cal V}(\varphi)$. This in
some sense is equal to introducing a creation field, see Ref.
\cite{HN}. Thus we have \be w+1= {p+\rho\over \rho}={-{\dot
\varphi}^2\over -{\dot \varphi}^2/2+{\cal V}(\varphi)}={2\over
3}\epsilon \rightarrow -\infty. \ee Due to ${\cal V(\varphi)}>
{\dot \varphi}^2/2$, this inequality actually requires the
denominator $\sim 0$, which gives ${\cal V(\varphi)}\sim {\dot
\varphi}^2/2$. Further, combining ${\dot \varphi}^2/2={\dot h}=
|\epsilon|h^2$, we may approximately deduce \be {\dot \varphi}^2/2
\simeq {\cal V}(\varphi)\simeq |\epsilon|h^2 .\label{phiv}\ee
This operation dose not mean that the phantom field is actually
required in the island universe model. The aim that we appeal to
the phantom field here is only to phenomenally simulate the NEC
violating fluctuation, which may be convenient to calculate the
primordial perturbation spectra from the NEC violating region. But
it maybe also a possibility that the phantom field only exists
instantly for producing the NEC violating fluctuation and after
the end of fluctuation, the phantom field thermalises into the
radiation, see Ref. \cite{FLP} for a different scenario.

In the momentum space, the equation of
motion of gauge invariant variable $u_k$, which is related to the
Bardeen potential $\Phi$ by $u_k\equiv a \Phi_k/\varphi^\prime $,
is \footnote{In the momentum space, the equation of motion of
gauge invariant variable $v_k$, which is related to the tensor
perturbation by $v_k \equiv a h_k$, is \be v_k^{\prime\prime}
+\left(k^2-{a^{\prime\prime}\over a}\right) v_k = 0 .\ee We have,
from (\ref{aeta}), $ a^{\prime\prime}/ a= (2-\epsilon )/
(\epsilon-1)^2 \eta^2$. In the regime $k\eta \rightarrow \infty $,
the mode $v_k \sim {e^{-ik\eta} \over (2k)^{1/2}}$ can be taken as
the initial condition.
In long wave limit, the expansion of the Bessel functions to the
leading term of $k$ gives \be k^{3/2}v_k /a\sim k^{-\nu+3/2}
\label{vk}\ee where $\nu =\sqrt{(2-\epsilon )/
(\epsilon-1)^2+1/4}$. } \be u_k^{\prime\prime}+\left(k^2
-{z^{\prime\prime}\over z}\right)u_k =0  .\label{uk}\ee We have,
from (\ref{aeta}), \be {z^{\prime\prime}\over z}=
{(1/a)^{\prime\prime}\over (1/a)} = {\epsilon\over
(\epsilon-1)^2\eta^2}\ee The general solutions of this equation
are the Bessels functions. In the regime $k\eta \rightarrow \infty
$, the mode $u_k$ are very deep in the horizon of $\Lambda$, see
Fig.1 and Fig.2. Thus
$ u_k \sim {e^{-ik\eta} \over (2k)^{3/2}}$ can be taken as the
initial condition.
In the
regime $k\eta \rightarrow 0$, the mode $u_k$ are far out the
horizon, and become unstable and grows. In long wave limit, the
expansion of the Bessel functions to the leading term of $k$ gives
\be
k^{3/2} u_k
\sim {\sqrt{\pi}\over 2^{3/2}\sin{(\pi
\nu)}\Gamma(1-\nu)}({-k\eta\over 2})^{-\nu+1/2} ,\label{kuk}\ee
where $\nu=\sqrt{\epsilon/(\epsilon-1)^2+ 1/4}$. We can see that
if $\nu\simeq 1/2$, the spectrum of $\Phi_k$ will be nearly scale
invariant. This requires \be |{\epsilon\over (\epsilon-1)^2} |\ll
1 .\label{spec}\ee Thus for the NEC violating fluctuation with
$\epsilon\rightarrow -\infty$ \footnote{We notice from
(\ref{spec}) that for $\epsilon\simeq 0_-$ in which the scale
factor expands exponentially, the spectrum is also nearly scale
invariant. This is the so called phantom inflation proposed and
discussed in detail in Ref. \cite{PZ1}, see also Refs. \cite{GJ,
NO, ABV, BFM}.
This result is actually also a direct reflection of the dualities
of primordial perturbation spectra, in which the spectrum index of
$\Phi$ is invariant under the change $\epsilon \rightarrow
1/\epsilon $, which has been pointed out in Refs. \cite{KST, BST}
and further studied in Refs. \cite{PZ2, Piao, Lid}, and see also
Refs. \cite{CL, DSS, Calca} for other discussions on the
dualities.}, we will have the nearly scale invariant spectrum
\cite{PZ}.

In $k\eta\rightarrow 0$, the Bardeen potential can also be
rewritten as \cite{MFB} $\Phi = {\cal C}+h{\cal D}/a, $ where
${\cal C}$ and ${\cal D}$ are constants dependent on the mode $k$.
In the inflation model in which $\epsilon\simeq 0$, $h$ is nearly
unchanged while $a$ increases exponentially, thus ${\cal D}$ is a
decay mode, and $\Phi$ is dominated by the constant mode ${\cal
C}$, and thus is nearly constant in superhorizon scale. This is
the reason why people can briefly obtain the amplitude of
perturbation reentering into the horizon by calculating the
amplitude of perturbation leaving the horizon during the
inflation. But for $\epsilon\rightarrow -\infty$, the case is just
inverse. $a$ is nearly unchanged while $h$ increases rapidly, thus
instead of being regarded as the decay mode, ${\cal D}$ is a
growing mode, which can be seen from (\ref{phik}), and compared
with the constant mode ${\cal C}$, it will dominate the Bardeen
potential $\Phi$, which means that the amplitude of perturbation
in the superhorizon scale will be still increasing during the NEC
violating fluctuation,
while
during the radiation/matter domination after the thermalisation,
${\cal D}$ will not increase any more but decrease, and thus the
constant mode ${\cal C}$ becomes dominated. Thus the key of
obtaining the scale invariant spectrum during the radiation/matter
domination is making the growing mode of $\Phi$ spectrum have an
opportunity to be inherited by the constant mode after the
thermalisation. However unfortunately, it has been shown \cite{BF}
that in the simple scenario the growing mode of $\Phi$ can hardly
be matched to the constant model after the transition. This result
is disappointing. However, it may be conceivable that our
simplified operation in the calculation of primordial spectrum may
have missed what. We here only phenomenally simulate the NEC
violating process by using the scale field violating the NEC, and
actually do not know the details of this NEC violating quantum
fluctuation and subsequent thermalisation. The latter may
significantly affect the final matching result.
The example that in the gauge, in which instead of occurring a
constant energy hypersurface the thermalisation may be everywhere
simultaneous, the growing mode of $\Phi$ spectrum can be inherit
by the constant mode at late time has been pointed out
\cite{DurrerV, GKST} \footnote{The recent numerical studies
\cite{C, BV} shown that the comoving curvature perturbation
$\zeta$ passes continuous through the transition and therefore the
spectrum inherited by the Bardeen potential at late times is the
one of $\zeta$ and not the spectrum of the growing mode of the
Bardeen potential. However, the examples (or classes) discussed in
these studies are still limited, thus whether the result is
universal remains open. }. We will tentatively take this
optimistic matching for the following calculations and leave
behind some further comments in the final conclusion. Because in
superhorizon scale the Bardeen potential is increased during the
NEC violating fluctuation and constant after the end of
fluctuation, it may be reasonable to take the value at the end of
fluctuation to calculate the amplitude of perturbation, \be
k^{3/2}\Phi_k\simeq {1\over 2^{3/2}}({\varphi^\prime \over
a})\simeq {\cal V}_e^{1/2}/2 \label{phik} .\ee where (\ref{phiv})
has been used.


The observations of CMB constrain the amplitude of perturbation is
$k^{3/2}\zeta_k\sim k^{3/2}\Phi_k\sim 10^{-5}$, which require
${\cal V}_e\sim 10^{-10} $, which corresponds to $h^2_e \sim
10^{-10}/|\epsilon|$. We assume the thermalisation is almost
instantaneous, thus $h_e^2\sim \rho_{r}$, where $\rho_{r}$ is the
energy density of radiation after the thermalisation. For example,
for $\rho_{r}\sim m_{ew}^4 \sim 10^{-60}$, where $m_{ew}$ is the
electroweak scale, which may be regarded as a loose lower limit of
reheating temperature though the lowest limit might be in
nucleosynthesis scale, we obtain $|\epsilon |\sim 10^{50}$. The
change of $a$ is far smaller than that of $h$, thus from
(\ref{caln}), we have ${\cal N}\simeq \ln{(h_e/h_i)}\sim
\ln{({\cal V}_e^{1/2}/|\epsilon|^{1/2}\Lambda^{1/2}})\sim
 127-\ln{|\epsilon|}/2\simeq 69$, which
is enough to solve the horizon problem of our observable universe
\cite{LL}. These results are also implicitly reflected in the
depiction of causal structure of island universe of Fig.2.  .

How do we distinguish the island universe from the inflation model
by the observations?
For $\epsilon\rightarrow -\infty$, from (\ref{vk}), we have
$\nu\simeq 1/2$, and thus the index of tensor perturbation is 2,
which means that the island universe model will produce a more
blue gravitational wave spectrum than the inflation model. This
character is also actually a reflection of rapid change of
background during the NEC violating fluctuation, compared with the
nearly unchanged background during the inflation. This result
generally leads to the intense suppression of gravitational wave
amplitude on large scale. Thus it seems for the island universe
model to be hardly possible to search for the imprint of
gravitational wave in the CMB, while a stochastic gravitational
wave will be consistent with the inflation model.

In conclusion, it seems to be possible that we can live in
``Islands", but slightly not optimistic, since the loophole how
the perturbations propagate through the thermalisation surface
remains, which may be fatal to the island universe model.
However, it should be mentioned that the NEC violating fluctuation
is actually a quantum phenomenon, and thus the innocence about the
thermalisation and matching hypersurface may to some extent
reflect the lack of our understanding on the quantum characters of
gravity. We think that
(at least) at present it will be premature to either confirm or
deny the island universe model. The deep insights into the quantum
gravity in the future may show us the final answer. However,
whatever the outcome, it appears that the viability of the island
universe model hangs in the balance and the different outcomes
will either hold out or end its competing as an alternative
scenario of the origin of observable universe.
We will back to the relevant issues in the future.

\textbf{Acknowledgments} The author would like to thank Sourish
Dutta, and especially Tanmay Vachaspatian for many valuable
comments of earlier draft and discussions. The author also very
thank Alexander Vilenkin for encouraging comments and Bo Feng,
Xinmin Zhang for helpful discussions. This work is supported in
part by K.C. Wang Postdoc Foundation, also in part by NNSFC under
Grant No: 10405029.

\end{document}